\documentclass{article}
\usepackage[utf8]{inputenc}
\usepackage[affil-it]{authblk}
\usepackage{float}

\title{Simple Model of a Standing Vertical Jump}
\author{Chris L. Lin}
\affil{Department of Physics, University of Houston, Houston, TX 77204-5005, United States of America.}
\date{\today}

\usepackage{graphicx}
\usepackage{comment}
\usepackage{amsmath}

\begin{document}
\noindent This is the Accepted Manuscript version of an article accepted for publication in European Journal of Physics. Neither the European Physical
Society nor IOP Publishing Ltd is responsible for any errors or omissions in this version of the manuscript or any version derived from it. The
Version of Record is available online at
https://doi.org/10.1088/1361-6404/abc85f.

{\let\newpage\relax\maketitle}

\setlength\parindent{0pt}
\section*{Abstract}
In this paper we use Newton's 3rd law to deduce the simplest model of a system that can perform a standing vertical jump -- a two-segmented object with an initial constant repulsive force between the segments, followed by an abrupt attractive force. Such an object, when placed on a sturdy ground, will jump, and the motion can be calculated using only the constant acceleration equations, making the example suitable for algebra-based physics. We then proceed to solve for the motion of an $n$-segmented object and determine the optimal number of segments for jumping. We then discuss a few similarities and differences of this simple model from jumping robots and jumping humans and conclude by arguing the model's pedagogical merits.

\section{Introduction}

Free fall occupies a large chunk of a standard course in introductory mechanics. For completeness, once Newton's laws are learned, the physics of the takeoff should be discussed. Although humans are in a privileged position of not having to jump to navigate or to escape predators, jumping still holds enchantment, an expression of joy, or a climactic flourish to a sequence of moves in sports and dance. One can leverage the popularity of sports \cite{goff2010gold,froh} and dance \cite{laws1984physics,doi:10.1119/1.4766448} to get students excited at applying the physics they learned to investigate the jump through construction of a simple model. Moreover, for students in the life and health sciences, biomechanics is one of the most important applications of physics \cite{winter2009biomechanics,knudson2007fundamentals} for their careers, and this exercise can serve to emphasize that physics governs both the animate as well as the inanimate.  \newline

Most discussions on jumping take as a starting point a force that develops between the object and the ground. Although a lot can be extracted from this model, particularly if a force plate is used to measure the ground force \cite{fplate1,foot,doi:10.1119/1.1976478,linthorne}, one drawback is that we know that jumping is initiated internally through muscle contractions before the force is communicated to the ground, so from a pedagogical perspective it would be nice to have a simple solveable model that attempts to show how internal forces causally lead to the development of external forces. Moreover, since this model considers the internal structure of the jumper, we will be able to predict the ground force generated by a human jumper, finding good agreement with force plate measurements.\newline

In section \ref{qsec2} we use Newton's 3rd law to deduce the simplest model of an object that can perform a standing vertical jump -- a two-segmented object with an initial constant repulsive force between the segments, followed by an abrupt attractive force -- and solve this model in section \ref{qsec3}. We then proceed to solve for the motion of an $n$-segmented object in section \ref{qsec4}, and in section \ref{interpretsec} determine the optimal number of segments for jumping. In section \ref{qsec6} we discuss similarities and differences of this simple model from jumping robots and jumping humans, and then conclude by arguing the model's pedagogical merits and broadly describing an experiment that students can perform.

\section{Deducing the model}\label{qsec2}

 \begin{figure}
\begin{center}
    \includegraphics[scale=.5]{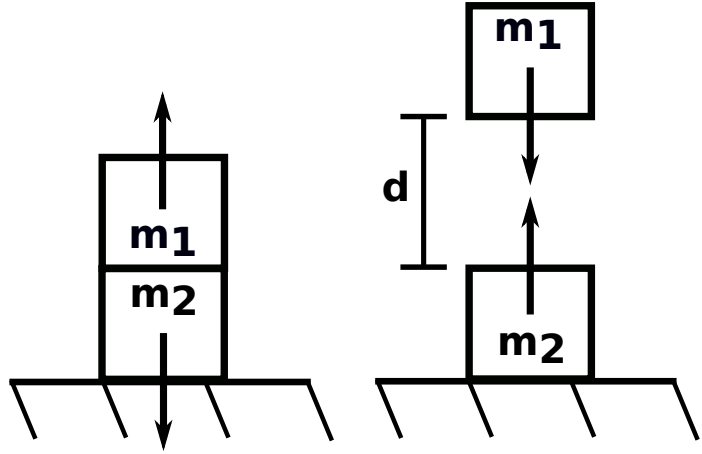}
\end{center}
\caption{Left -- an internal repulsion causes separation of the two segments. Right -- after a max separation distance $d$, the opposite force-pair become attractive.}
\label{fig1}
\end{figure}

Consider an organism made of two parts, a top segment of mass $m_1$, and a bottom segment of mass $m_2$, resting on a hard surface. Given that internal forces between the top and bottom segments must be equal but opposite, one can ask students whether the internal force pair that initiates the jump is attractive or repulsive. Students can reason that the top segment must move upwards before the bottom segment can, so there must be an upward force produced internally on the top segment, and from Newton's 3rd law, an equal downwards force on the bottom segment, hence initially there is repulsion: see Fig. \ref{fig1}.
 The top segment accelerates upward, but the bottom segment is prevented from accelerating downward due to the ground, which therefore supplies an external upwards force. As the top segment displaces upward, its separation from the bottom segment (which remains static, pressed into the ground) increases. This separation cannot continue indefinitely if the object is to not break apart, so the internal forces must change their direction and become attractive as the upper segment pulls the bottom segment upwards, and from Newton's 3rd law, is itself pulled downwards by the bottom segment. One can then ask students to make an analogy of this model with jumping by humans, and would likely get answers discussing how, from a crouched position, muscles exert internal forces causing humans to unfold, but once the human has completely straightened to reach maximum separation, muscle forces switch to joint forces to prevent the two halves from separating.

\section{Solving the model}\label{qsec3}

To calculate the speed at which the object leaves the ground, we model the initial phase of the jump as comprising a constant repulsive force $F$ exerted over a maximum separation distance of $d$ between the two segments. Once the distance $d$ is reached, we model the attractive force between the two segments as an abrupt completely inelastic collision, after which the segments are locked together. \\

The velocity of the upper segment under constant acceleration $a=\frac{F-m_1 g}{m_1}$ over a distance $d$ is

\begin{equation}
v_f=\sqrt{2\left( \frac{F-m_1 g}{m_1}\right)d+v_i^2 }.
\end{equation}

We model the second phase as a completely inelastic collision between the lower and upper segments, where the final velocity of the combined system is:

\begin{align} \label{eq2}
v'_f&=\frac{m_1 v_f}{m_1+m_2}\\ \nonumber
&=\frac{m_1}{m_1+m_2}\sqrt{2\left( \frac{F-m_1 g}{m_1}\right)d +v_i^2}.
\end{align}

Once in the air, the object is in free-fall and the height of the jump is given by

\begin{align}\label{jumpheight12}
h&=\frac{v'^2_f}{2g}\\ \nonumber
&=\frac{m_1}{(m_1+m_2)^2}\left(\frac{F-m_1g}{g}\right)d,
\end{align}

where $v_i$ was set to zero because the upper segment is initially at rest\footnote{If there is some initial spacing between the top and bottom segments, the possibility exists for $v_i<0$ (while still retaining $F>0$) which is known as a countermovement jump. In a countermovement jump, the muscles undergo eccentric contraction. We do not consider countermovement jumps in this paper.}. The height $h$ is defined as the distance from the ground to the bottom of the feet when the object has reached its maximum height in the air.\\

Traditionally, collisions are treated long after Newton's 3rd law and the constant acceleration equations \cite{serway}. We therefore in Appendix \ref{impulseapprox} offer an alternate derivation of Eq.\eqref{eq2} using only the constant acceleration equations so that the model can be introduced earlier. We note that the forces in this model qualitatively capture what would happen if the two masses were connected by a spring and the top mass depressed then released: there would be repulsion as the top mass accelerated upwards, with attraction commencing when the spring reached its resting length and growing in strength thereafter, and takeoff would occur when the spring moved past its resting length by an amount equal to the weight of the bottom mass divided by the stiffness of the spring. \newline

We end this section by giving a rough estimate for the parameters for this two-segment model, while in the next section we consider the addition of more segments. A jump using only two segments is often seen when blocking in volleyball, where the knees are bent but the upper body is fairly straight, so that the top segment is all parts of the body above the knee (three segments are needed when the upper-body is folded forward at the hip: see Fig. \ref{figJump13}). The average person's lower leg and feet comprise approximately $1/10$ the total mass $m_T$ of the person \cite{segmentdata}, so that $m_1=0.9 \,m_T \,g$ and $m_2=0.1 \,m_T \,g$. For the distance $d$ we will assume that the upper and lower legs are both $0.5$ meters long and are initially bent at 90 degrees to each other, so that they unfold through an approximate distance of $d \approx 0.3$ meters. The parameter $F$ is the most difficult to determine, but as a rough estimate we can use the result of section \ref{interpretsec} where $F \approx \frac{5.7}{3}\, m_T\, g$. Plugging these values into Eq. \eqref{jumpheight12}, we get a jump height of around $h \approx 0.3$ meters. 

\section{Extension to N-segments}\label{qsec4}

\begin{figure}
\begin{center}
    \includegraphics[scale=.25]{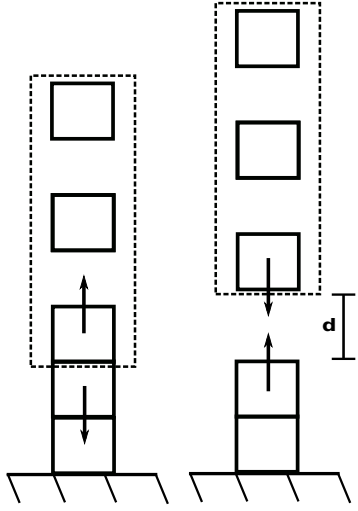}
\end{center}
\caption{The multi-segment system treated like a two-segment system, where one of the segments is the dotted box comprising n-1 segments, and the $n^{\text{th}}$ segment lies just below.}
\label{fig2}
\end{figure}

We simplify the problem by having all segments be the same, $m_1=m_2=...=m_{N}=m$, and that the $n^{\text{th}}$ segment does not turn on its repulsion until the $(n-1)^{\text{th}}$ segment has collided with the $(n-2)^{\text{th}}$ segment: see Fig. \ref{fig2}. At that moment denote the velocity of the entire upper mass of $(n-1)m$ as $v_{n-1}$. After completing the interaction with the $n^{\text{th}}$ mass, the new velocity $v_{n}$ of the combined mass $nm$ is, using Eq. \eqref{eq2},

\begin{equation}\label{eq4}
v_{n}=\frac{(n-1)m}{(n-1)m+m}\sqrt{2\left( \frac{F-(n-1)m g}{(n-1)m}\right)d +v_{n-1}^2}\,.
\end{equation}

Plugging in a few numbers

\begin{align}
&v_2=\frac{1}{2}\sqrt{2\left( \frac{F-m g}{m}\right)d +0^2}=\sqrt{\frac{1}{2}}\sqrt{\frac{d}{m}}\sqrt{F-mg}\\ \nonumber
&v_3=\frac{2}{3}\sqrt{2\left( \frac{F-2m g}{2m}\right)d +v_2^2}=\sqrt{\frac{2}{3}}\sqrt{\frac{d}{m}}\sqrt{F-\frac{5}{3} mg}\\ \nonumber
&v_4=\frac{3}{4}\sqrt{2\left( \frac{F-3m g}{3m}\right)d +v_3^2}=\sqrt{\frac{3}{4}}\sqrt{\frac{d}{m}}\sqrt{F-\frac{7}{3} mg}.
\end{align}

From the pattern we guess $v_n=\sqrt{\frac{n-1}{n}}\sqrt{\frac{d}{m}}\sqrt{F-\frac{2n-1}{3} mg}$ which we can verify by indeed showing that it satisfies Eq. \eqref{eq4}. Therefore for $N$ identical segments the velocity upon takeoff and maximum height attained are:

\begin{align}\label{heightAttained}
v_f&=\sqrt{\frac{N-1}{N}}\sqrt{\frac{d}{m}}\sqrt{F-\frac{2N-1}{3} mg}\\ \nonumber
h&=\frac{v_f^2}{2g}=\frac{N-1}{N}\left(\frac{1}{2m}\right)\left(\frac{F-\frac{2N-1}{3}mg}{g}\right)d.
\end{align}

Eq. \eqref{heightAttained} is also derived in Appendix \ref{pseudowork} using the work-kinetic energy theorem. 

\section{Interpreting the N-Segment result}\label{interpretsec}

If we express the force $F$ between segments as a multiple $\alpha$ of the weight $mg$ of each segment ($F=\alpha\, m g$, so that $\alpha$ is the strength-to-weight ratio of each segment), then the height in Eq. \eqref{heightAttained} becomes:

\begin{align} \label{xHeight}
h=\frac{1}{2}\left(\frac{N-1}{N}\right)\left(\alpha+\frac{1}{3}-\frac{2N}{3}\right)d.
\end{align}

\begin{figure}[h]
\begin{center}
    \includegraphics[scale=.75]{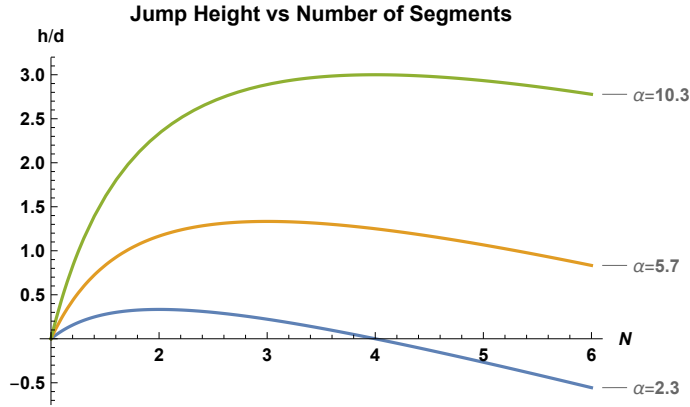}
\end{center}
\caption{Jump height ratio $h/d$ vs number of segments $N$, for different values of the strength-to-weight ratio $\alpha$ of each segment. The curves have a decreasing tail that goes as $h/d=\frac{\alpha+1}{2}-\frac{N}{3}+O[\frac{1}{N}]$, which is the large-$N$ limit of Eq. \eqref{xHeight}.}
\label{figSephiroph}
\end{figure}

The height of the jump is proportional to the uncoiling length $d$ between segments, but for a given $\alpha$, different values of $N$ maximize the height $h$. Students in algebra-based physics courses can numerically plug in different $\alpha$ values and use graphing tools to find the $N$ value which maximizes $\frac{h}{d}$: see Fig. \ref{figSephiroph}.\\

Calculus yields the exact result of $N_{\text{max}}(\alpha)=\sqrt{\frac{3\alpha+1}{2}}$ and $h_\text{max}(\alpha)=h(N_{\text{max}}(\alpha))=\frac{(\sqrt{3\alpha+1}-\sqrt{2})^2}{6}d$, \footnote{One can avoid calculus by noting that Eq. \eqref{xHeight} can be written as $h=\frac{d}{2}(\alpha+1)-\frac{d}{3}\left(N+\frac{k}{N}\right)$, where $k=\frac{3\alpha+1}{2}$. $\left(N+\frac{k}{N}\right)$ is minimized by noting $\left(N+\frac{k}{N}\right)^2=\left(N-\frac{k}{N}\right)^2+4k$ which has a minimum when $N=\frac{k}{N}$, or $N=\sqrt{k}=\sqrt{\frac{3\alpha+1}{2}}$.} which implies that the greater the force each segment can exert relative to its weight, the more segments you
can add to maximize your jump before adding becomes counterproductive. When $\alpha=2.3$ and $\alpha=5.7$ the maximum height occurs when there are two and three segments (respectively).\\

For humans, the number of segments can be estimated to be between 2 and 3 (e.g. one can uncoil at both the knees and the hips using the quadriceps and glutes, respectively), suggesting $2.3 <\alpha<5.7$. In Eq. \eqref{groundforce} we show that the ground force (as would be measured by a force plate) throughout the jump varies from a minimum value of $F^\text{min}_\text{g}=(\alpha+1)mg$ to a maximum value of $F^\text{max}_\text{g}=(\alpha+N-1)mg$, which when written in terms of the total mass $M=N m$ corresponds to $F^\text{min}_\text{g}=\frac{(\alpha+1)}{N}Mg$ and $F^\text{max}_\text{g}=\frac{(\alpha+N-1)}{N}Mg$. Plugging in $\alpha=2.3$ for $N=2$ and $\alpha=5.7$ for $N=3$ predicts that a force plate would measure $1.7 Mg$ for 2 segments and between  $2.2 Mg$ and $2.6 Mg$ for 3 segments. This is not too far from force plate measurements which put the the force measured as between $2 Mg$ and $2.4 Mg$ \cite{doi:10.1119/1.1976478}.

\section{Mechanical realization of multiple segments}\label{qsec6}

\begin{figure}
\begin{center}
    \includegraphics[scale=.33]{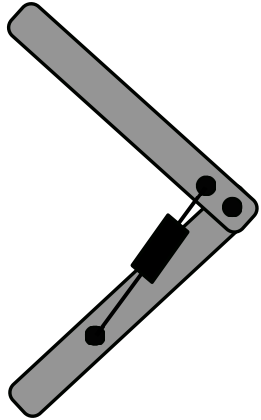}
\end{center}
\caption{Two segments that repel via the hydraulic cylinder.}
\label{fig3}
\end{figure}

The two-segment model kind of resembles a robot leg powered by a hydraulic cylinder, where expansion of the cylinder provides the repulsive force and separation of the upper and lower segments is achieved by opening at the hinge: see Fig. \ref{fig3}. It should be noted that the cylinder, being connected to both the lower and upper segments, must deform as the two segments separate, which it does by extending. \newline

\begin{figure}
\begin{center}
    \includegraphics[scale=.33]{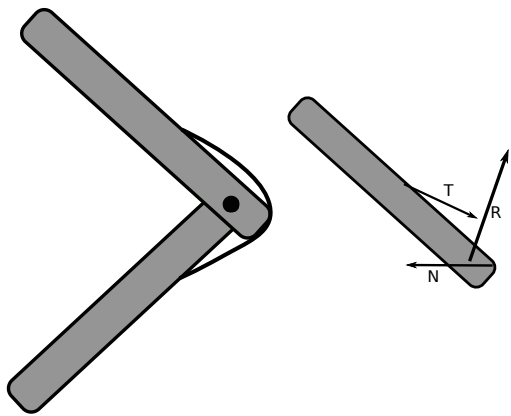}
\end{center}
\caption{Two segments that repel via the muscle-joint-kneecap system, the sum of whose forces has an upward component.}
\label{fig4}
\end{figure}

Muscle and joints are more complicated. In Fig. \ref{fig4}, the contraction of the muscle, which we model with a tension $T$ in the string, causes an attractive rather than a repulsive force. However, the hinge itself provides a reaction force $R$ that ultimately causes a net upward force on the upper segment. The force $N$ comes from the normal force of the string wrapped around the kneecap. Analysis of such a device is most natural in terms of torque, where it becomes obvious that the tension $T$ causes the center of mass of the top segment to move upwards as it rotates clockwise, despite the fact that it is pulling downward.\newline

We note that attraction from the joint to prevent separation of the segments acts at each instant in time and does not first wait for the repulsive action to fully complete. The lower leg is pulled upward (counterclockwise in Fig. \ref{fig3}) and clears the ground when it is sufficiently vertical that the foot no longer rotates into the ground due to the torque from the joint.

\section{Conclusion}

\begin{figure}[H]
\begin{center}
    \includegraphics[scale=.7]{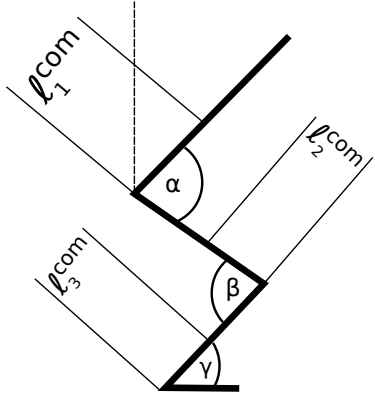}
\end{center}
\caption{The jumping motion is approximated by the top segment rotating to the dashed vertical line, followed by $\beta$ opening to 180 degrees during which the top segment remains vertical and the feet stationary. The length of the segments, from top to bottom, are $\ell_1$, $\ell_2$, and $\ell_3$.}
\label{figJump13}
\end{figure}

 Accurate biomechanical models of humans can go as far as having 16 rigid bodies joined together comprising 38 degrees of freedom \cite{Herr467}, which resists any attempts at an analytic solution. Analytic solutions for the jumping motion of springs \cite{hoop,springbok} have been made, but as the forces are not constant, they require the solving of differential equations. The simple, torqueless, algebraic model presented here can serve as a budget model against which more sophisticated models \cite{10.230755945} can be compared. Equally important in getting students to think about what a model contains is to get them to consider what it lacks. A rather direct way to do this is to have them to test the model against the real world, which will also give students practice in measuring and curve fitting. We give a rough sketch of how this might go. We first solve the 3-segment model but with differing masses $m_1$, $m_2$, $m_3$; distances $d_{1}$, $d_{2}$; and forces $F_1$, $F_2$. $d_1$ is the displacement of the center of mass of $m_1$, $d_2$ the displacement of the center of mass of $m_1$ and $m_2$ when they are treated as a combined system, $F_1$ the force between $m_1$ and $m_2$, and $F_2$ the force between $m_2$ and $m_3$. To translate the parameters of a real jump into parameters of the model, we make the measurements indicated in Fig. \ref{figJump13}. \\
 
 The upper-body first rotates through an angle of $90-\alpha+\beta-\gamma$ to reach the vertical dashed line during which $\beta$ and $\gamma$ remain constant. We treat this as $m_1$ whose center of mass moves through a vertical distance  $d_1=\ell_1^{\text{com}} (1-\cos(90-\alpha+\beta-\gamma))$ under a force $F_1$. Then with the upper-body remaining vertical, $\beta$ opens to 180 degrees while the feet remain fixed. We represent this as the center of mass of a combined $m_1$ and $m_2$ moving through a vertical distance of $d_2=\frac{m_1\Delta y_1+m_2\Delta y_2}{m_1+m_2}$, where $\Delta y_1=(\ell_2+\ell_3)-(\ell_2\sin(\beta-\gamma)+\ell_3\sin(\gamma))$, $\Delta y_2=(\ell_3+\ell_2^{\text{com}})-(\ell_3 \sin (\gamma)+\ell_2^{\text{com}}\sin(\beta-\gamma))$, under a force $F_2$. $F_2$ can be determined by measuring the jump height when $\ell_1$ starts at (and stays) vertical, essentially using only the knees (quad-dominant) for the jump. With $F_2$ determined, $F_1$ can be determined by having $\alpha$ be very small (glute-dominant) and measuring the height of the jump along with the previously measured value of $F_1$. The 3-segment model then has prediction for the height that the test subject can achieve as a function of any initial $(\alpha, \beta,\gamma)$\footnote{If the foot did not exist (so that the lower leg terminated at a point), then $\gamma=\gamma(\alpha,\beta)$, i.e., the body would adjust $\gamma$ for a given hip $\alpha$ and knee $\beta$ angle so that the student's center of mass remained above the terminal point. Due to the length of the foot, $\gamma$ is allowed some freedom and is therefore treated as an independent initial parameter in the model.}. One can then ask students whether the experiment was successful, and if not, ask for possible reasons for its failure, which can include difficulties in experimental measurement, dubious assumptions about the human body, or even incorrect simplifications of the physics. An example could be that the force generated by the muscles depend on the angle ($F_1=F_1(\alpha)$, $F_2=F_2(\beta)$) or even the rate of change of the angle, or that the approximation for the sequence of movements described here is too crude. It should be noted that regardless of the precise sequence of movements involved in jumping, starting from the initial position indicated in Fig. \ref{figJump13}, the total center of mass will have moved through a distance $\Delta y_{\text{com}}=\frac{m_1\Delta y_1+m_2\Delta y_2+m_3\Delta y_3}{m_1+m_2+m_3}$, where $\Delta y_3=\ell_3^{\text{com}}(1-\sin(\gamma))$, $\Delta y_2=\ell_3(1-\sin(\gamma))+\ell_2^{\text{com}}(1-\sin(\beta-\gamma))$, and $\Delta y_1=\ell_3(1-\sin(\gamma))+\ell_2(1-\sin(\beta-\gamma))+\ell_1^{\text{com}}(1-\sin(\alpha-(\beta-\gamma)))$. One can then use this value in models where only the total center of mass displacement is considered (as opposed to considering individual hip and knee angles).\\

 \textbf{\large Acknowledgements}\\
 
 The author would like to acknowledge the anonymous reviewers for their helpful suggestions.


\appendix

\section{Constant force impulse approximation}\label{impulseapprox}
To derive Eq. \eqref{eq2} without collision equations, we assume that after the repulsive force $F$ acts for a distance $d$, the constant attractive force $F'$ acts for a distance $\epsilon$, after which both segments lock together and move at the same velocity $v'_f$. The top segment deaccelerates to 

\begin{align} \label{eq9}
    v'_f=\sqrt{v^2_f-2\frac{F'}{m_1}\epsilon}
\end{align}

in time $t=\frac{v_f-v'_f}{\frac{F'}{m_1}}$, while the bottom segment accelerates to 

\begin{align} \label{eq10}
    v'_f&=\frac{F'}{m_2}\left(\frac{v_f-v'_f}{\frac{F'}{m_1}} \right)\\ \nonumber
    v'_f&=\frac{m_1 v_f}{m_1+m_2},
\end{align}

which proves Eq. \eqref{eq2}. Moreover, we see that for Eqs \eqref{eq9} and \eqref{eq10} to be equal, we must have

\begin{align} \label{eq11}
    F'=\left(\frac{v^2_f}{2\epsilon}\right)\frac{m_1 m_2(2m_1+m_2)}{(m_1+m_2)^2}.
\end{align}

Therefore we have the freedom to make $\epsilon$ very small with the corresponding $F'$ in Eq. \eqref{eq11} very large, such that their product in Eq. \eqref{eq9} is unchanged. We can then neglect $\epsilon$ whenever compared to $d$, so that we may still say that two segments lock and move together when their separation distance reaches $d$.

\section{N-segments using work kinetic-energy Equation}\label{pseudowork}

When $n$ segments are in the air, the ground must support the weight of $N-n$ segments, along with the force $F$ pressing on these segments:

\begin{equation} \label{groundforce}
    F^{(n)}_{\text{ground}}=(N-n)mg+F.
\end{equation}

Therefore the net external force on the entire system when $n$ segments are in the air is: 

\begin{equation}
    F^{(n)}_{\text{net}}=F^{(n)}_{\text{ground}}-Nmg=F-nmg.
\end{equation}

When $n$ segments move upwards a distance $d$, the center of mass of the entire system moves upward by:

\begin{equation}
\Delta X^{(n)}_{\text{COM}}=\frac{nd}{N}.
\end{equation}

Therefore using the work kinetic-energy equation \cite{doi:10.1119/1.1845983,doi:10.1119/1.2909743,sherwood}:

\begin{align}
\sum_{n=1}^{N-1} F^{(n)}_{\text{net}}\Delta X^{(n)}_{\text{COM}}&=\Delta \text{K}\\ \nonumber
\sum_{n=1}^{N-1} \left(F-nmg\right)\frac{nd}{N}&= \text{K}_\text{f}.
\end{align}

Using the famous sums $\sum_{n=1}^{N-1}n=\frac{N(N-1)}{2}$, $\sum_{n=1}^{N-1}n^2=\frac{(N-1)N(2N-1)}{6}$, along with setting $K_f=Nmgh$ for free-fall, one gets:

\begin{align}
\left(\frac{N(N-1)}{2}F -\frac{(N-1)N(2N-1)}{6} mg \right)\frac{d}{N}&=N mgh\\ \nonumber
\frac{N-1}{N}\left(\frac{1}{2m}\right)\left(\frac{F -\frac{(2N-1)}{3} mg}{g} \right)d&=h,
\end{align}

which agrees with Eq. \eqref{heightAttained}.


\end{document}